\def\ap{$\sim$}

\def\etal{{\it et al.}}
\def\g{$\gamma$-ray}
\def\co{{\it Compton Observatory}}

\def\pt#1{$10^{#1}$}
\def\tpt#1{$\times10^{#1}$}

\documentclass[11pt,preprint2]{aastex}
\begin{document}
\onecolumn
\Large

{\small
To appear in the Proceedings of the OJ-94 Annual Meeting 1999 
``Blazar monitoring towards the third
millennium'', held in Torino, Italy,  May 19--21, 1999
}
\medskip
\begin{center}
{\bf The Automatic Telescope Network}
\end{center}

\medskip
\normalsize
\begin{center}
{\bf 
J.R. Mattox
}

\medskip
{\small
{\sl 
Institute for Astrophysical Research, Boston University; 
Email: Mattox@bu.edu\\
} 
}
\end{center}

\bigskip\noindent
{\bf Abstract }
\vskip .6cm

Because of the scheduled  GLAST mission by NASA, there is strong 
scientific justification for preparation for very extensive blazar 
monitoring in the optical bands to exploit the opportunity to learn 
about blazars through the correlation of variability of the gamma-ray 
flux with flux at lower frequencies. Current optical facilities do 
not provide the required capability.Developments in technology have 
enabled astronomers to readily deploy automatic telescopes. The 
effort to create an Automatic Telescope Network (ATN) for blazar 
monitoring in the GLAST era is described.

\bigskip\noindent
{\bf 1. Introduction}
\vskip .6cm

The EGRET \g\ telescope aboard the \co\ has detected \ap1 GeV emission from \ap70 blazars (Hartman \etal\ 1999, Mattox 1999a). 
The  apparent \g\ luminosity seen for the EGRET blazars
is as much as one hundred times 
larger than that at
all other wavelengths for some flaring EGRET blazars. 
Variability of the \g\ flux
from some blazars on a time-scale as short as 4 hours has been observed
(Mattox \etal\ 1997). This implies that the region of \g\ emission
must be very compact.
Because the opacity for \g\ to
$\gamma-\gamma$ pair production with x-rays must not prevent 
$\gamma$-rays from
escaping, relativistic beaming with a Lorentz factor of \ap10 for the 
bulk of material in the jet is required (Mattox \etal\ 1997).
This conclusion is
reinforced by the observation of a high x-ray state during the
1996 \g\ flare of 3C 279 (Wehrle \etal\ 1998) which implies that the
x-rays originate in the same volume as the \g s). 
With substantial accretion onto a \ap\pt8 M$_\odot$ blackhole,
there is sufficient power to create the relativistic jets. However, the physics
involved in the conversion of gravitational potential to kinetic luminosity
is not  understood.

It is widely believed that this GeV emission  is due to
inverse-Compton scattering by shock-accelerated leptons within the relativistic jet.
However, there is  disagreement over the origin of the \ap1 eV photons which are
scattered. Some modelers believe that they originate in the
synchrotron emission of the leptons, so the
\g s are a result of the synchrotron self-Compton (SSC) 
process  (Bloom \& Marscher 1993).

Another possibility is that the low energy photons come from outside of the jet.
This is designated as the external Compton scattering (ECS) process.
Dermer, Schlickeiser, \& Mastichiadis (1992) suggested that 
they come directly from an accretion disk around a 
black hole at the base of the jet. It was subsequently  
proposed that the dominant source
of the low energy photons for scattering could be  due to
re-processing of disk emission
by broad emission line clouds   
(Sikora, Begelman, \&  Rees 1994).

The correlation of optical, x-ray, and \g\ emission is expected to provide
a definitive test of these models.
If the variation in the synchrotron
flux is due to a change in the electron density, the SSC emission, which depends
on the second power of electron density, will 
 be observed to vary quadratically in comparison to the synchrotron
with  no lag (for a single homogeneous emission zone).
If the high-energy emission is ECS due to reprocessed in the broad 
line region before inverse Compton scattering
(Sikora Begelman and Rees 1994), the
ECS flux will lag the optical disk flux by at least 1 day if the ECS flare
is due to increased optical emission from the disk. 
If the ECS flare is
due to an enhancement in the relativistic particle content of the jet,
the optical and \g\ flux will vary in a linear fashion with no lag if
the optical synchrotron and ECS emission occur in the same region of
the jet.
Ghisellini \& Madau (1996) suggest that the dominant source of low energy photons for ECS scattering
is broad-line-region re-processing of jet synchrotron  emission ---
the ``mirror model''. In this case, there could be linear correlation between
synchrotron  emission and ECS emission with a \g\ lag of \ap2\tpt3 seconds,
if jet plasma entering a region of enhanced magnetic field resulting
in more synchrotron  emission, and the re-processing region was \ap\pt3
light-seconds away. In this scenario, a change in the synchrotron  emission
spectrum would occur, and could be discerned with accurate photometry
in multiple optical bands (e.g., B and R).
Thus, good sampling of multiple energy ranges offers the opportunity to 
distinguish models for 
the continuum emission of blazar jets.

\begin{figure}
\epsscale{0.7}
\plotone{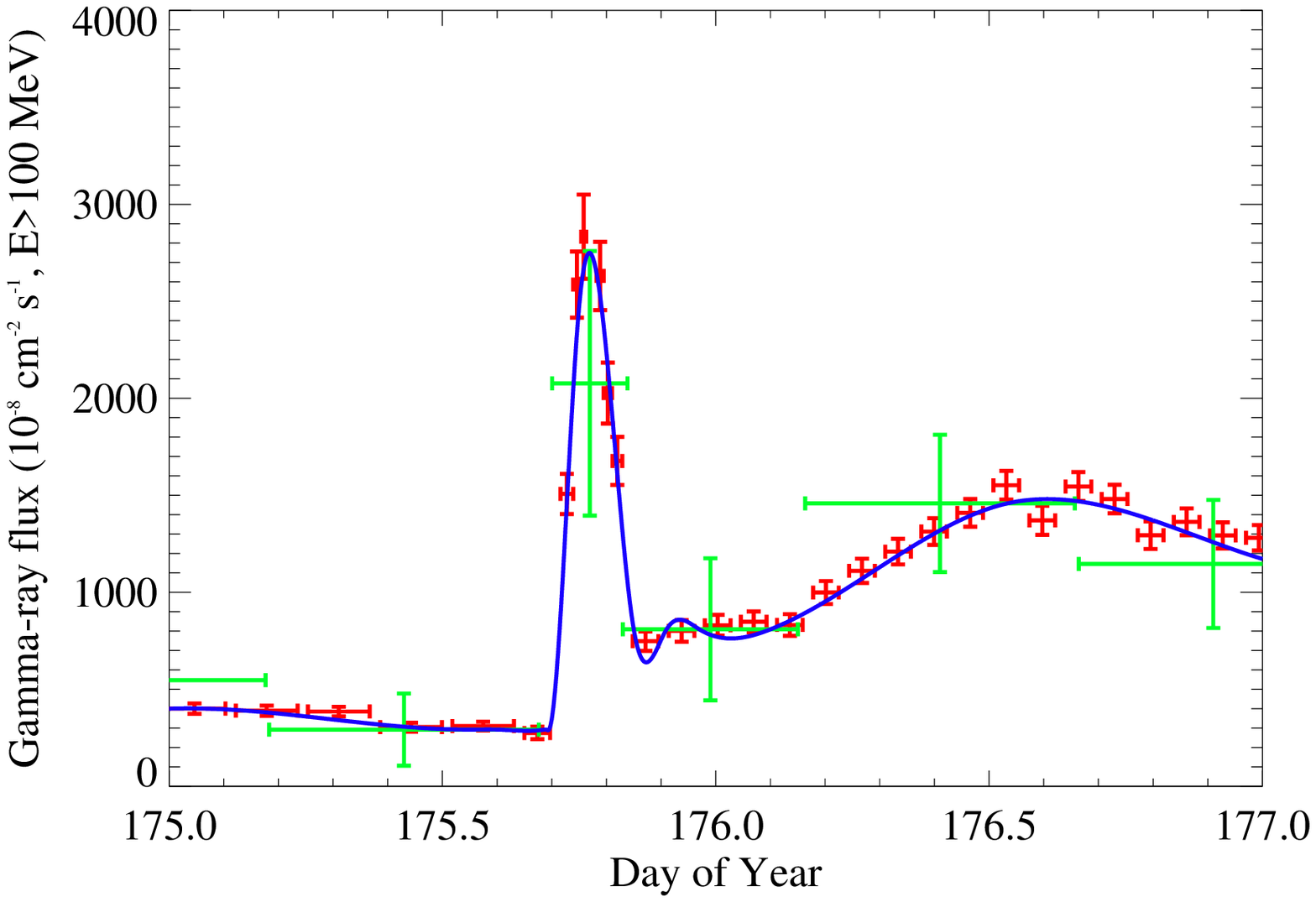}
\plotone{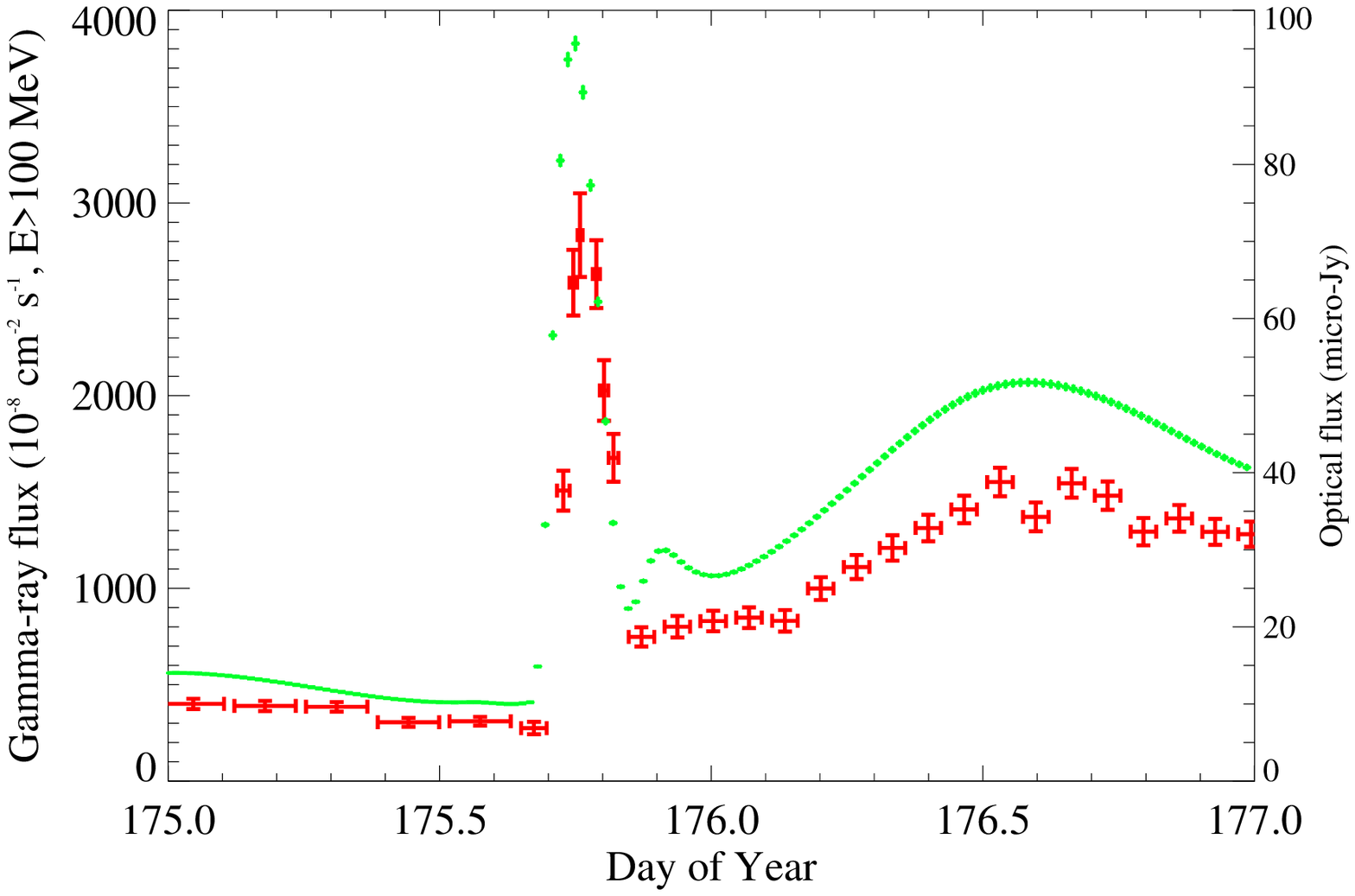}
\caption{The figure on top shows the measured EGRET flux for
PKS~1622-297 (Mattox \etal\ 1997)
with the large error bars [green in a color rendition].
The solid [blue] line is a plausible hypothesis for the actual flux. Assuming this
flux,
simulated GLAST measurements are shown with much smaller error bars [red].
The figure on the bottom shows the simulated GLAST measurements [red] along with
simulated  ATN optical observations [green].
It is assumed that the
\g\ flux  varies linearly in relationship to the optical with a lag
of 2000 seconds. 
The optical observation interval is 30 minutes. 
}
\end{figure}

As we gain understanding of this
emission, it potentially can then be used to study the origin of the jet.
In addition to providing for discrimination between models, simultaneous
multiwavelength observations also have the potential to determine
of properties of the jet; e.g., Takahashi \etal\ (1996) infer the strength of the
magnetic field in the jet of blazar Mrk 421 by examining the rate of Synchrotron
energy losses with the ASCA satellite.

NASA's GLAST mission, the next generation GeV gamma-ray 
telescope, will provide excellent
GeV sensitivity. It is currently scheduled to be launched in 2005 and
to operate for a minimum of 5 years.
NASA's URL for GLAST is http://glast.gsfc.nasa.gov.
A simulated blazar GLAST light curve is shown in Figure 1
for PKS~1622-297 (which produced the largest point source flux
observed by EGRET, Mattox \etal\ 1997). Compared to EGRET,
a dramatic enhancement in the \g\ light curve resolution due to the
increased aperture of GLAST is apparent.

\vskip .5cm
{\bf 2. The Optical Monitoring of Blazars in the GLAST Era}
\vskip .3cm

If the \g\ luminosity function of blazars is roughly similar to the
radio luminosity function of the parent population of
extragalactic flat-spectrum radio sources as often assumed (e.g., Stecker and Salamon 1996),
GLAST will detect \ap5000 blazars. To observe
just 20\% of this population in the optical band just
once per month will require 
33 observations per day. For the expected magnitude distribution
(\ap14 to \ap24 --- deduced from the distribution of the V magnitudes given by
NED for  robust
3EG EGRET identifications, Mattox 1999a),
a telescope of at least \ap1m aperture will be required for many of these
sources.
Assuming an average of 10 minutes to set up, expose, and read-out for each
observation, 80\%
of all available time with a 1m class telescope will be required for just
monthly monitoring. 

Since \g\ variability was observed with EGRET on sub-day
time scales, (see Figure 1), 
more frequent monitoring is appropriate for bright sources which will produce 
sufficient GLAST counts to be more rapidly resolved. 
Table 1 specifies the required
density of optical monitoring based 
on the number of  sources  which are expected to be time resolved by GLAST
on a timescale
shorter than \ap10 days.
The average number of optical observations is 222 per day, and requires four
1m class telescopes at a variety of longitudes. 
The prospect of making 
an average of  222 optical observations per day for the 
duration of the GLAST mission 
(4\tpt5 observations over 5 years)
provides a strong incentive to consider the use of automated,
ground-based telescopes.

This monitoring
could be accomplished from space by a single facility with capabilities
similar to the Hubble Space Telescope. However, the construction of four 
1m class telescopes on
the ground can be accomplished for a cost  \ap2 orders of 
magnitude less than a space telescope with HST capabilities.

\begin{table}
\begin{center}
\begin{tabular}{|cccccc|} \hline
Obs. interval& Obs. duration&Fraction & Number of sources &Obs./day&Telescopes required\\ \hline
Monthly & 10 min. & 0.2 &1000 & 33 & 0.8\\
Weekly & 10 min. &  0.05 & 250 & 36 & 0.9\\
Daily & 7 min. &  0.01 &50 & 50 & 0.8\\
2 hours & 5 min. &  0.001 & 5 & 60 & 0.7\\
5 minutes& 5 min. & 0.00003 & one 15\% of the time&288 (ave. 43) &0.5\\
Total & & & & 222 & 3.7\\
 \hline
\end{tabular}
\caption{
\baselineskip=11pt
Optical monitoring required during the GLAST mission. An average of 7 clear 
dark hours for each \ap1m telescope is assumed for each diurnal interval.
A measurement in two optical bands (i.e., B and R) could routinely be made 
for sufficiently bright sources in the specified observation interval
to ascertain the spectral slope.
}
\end{center}
\end{table}

In addition to the \g\ flux, Figure 1 shows 
simulated high-temporal-density
optical observations. It was assumed that the
\g\ flux is varies linearly in relationship to the optical with a lag
of 2000 seconds, consistent with the ``mirror model'' for \g\ emission.
The linear dependence and the lag are both clearly resolved.
Thus, we can expect that dense optical sampling in conjunction with 
GLAST observations
will produce a very good opportunity to test blazar models in detail.
Because PKS~1622-297 is a 20th magnitude source, a 
network of \ap1m
telescopes would be required to obtain this data.
Although a large multiwavelength campaign  was organized for 3C 279 in 1996
(Wehrle \etal\ 1998) which
explicitly included over 20 optical observers,
only one optical observation was obtained 
in the 3 days of maximum \g\ flux.
Without extensive preparation prior to the GLAST mission, more extensive
optical coverage may not be available during the GLAST mission.

\vskip 1cm
{\bf  3. Automatic Telescopes}
\vskip .3cm

The automatic operation of optical telescopes  began with a
 50-inch telescope on Kitt Peak.
This telescope was constructed with NASA funding as the
Remotely Controlled Telescope (RCT).
The initial intention was to develop techniques
for controlling telescopes in space. It was soon apparent that this was not
a useful approach to learning to control a space telescope 
--- the dynamics were very different, as were the scales of
the budgets (personal communication, Steve Maran, 1999). 
The RCT telescope focus then shifted to an attempt to demonstrate
the operation of an automated telescope --- something which the
Whitford Committee  suggested  in 1964 as a means to 
enhance the productivity of small telescopes (Maran 1967).
A decade of effort resulted in one astronomical paper (Hudson \etal\ 1971), and
the realization that a human telescope operator was much more cost effective than
telescope automation with the technology available in 1969 
(personal communication, S. Maran, 1999).

During the  3 decades which have transpired
since the  RCT telescope experiment,
remarkable advances in technology have occured.
Modern technology now makes telescope automation straight forward.
It is likely that  vision of the Whitford Committee of substantial gains in
productivity through the automation of telescopes may soon be realized.
The most important technological advances have been: (1) the development
of powerful, reliable, inexpensive, and compact computers;
(2) the development of intelligent
controllers for mechanical motions; (3) the development of
charge coupled devices (CCDs),
and (4) the accumulation of
experience in the most effective ways to control and use fully
automated and unattended observatories.

A telescope equipped with servo motors and rotation encoders on both axes, 
and driven by a computer with an accurate model of telescope flexure 
and pointing aberrations, can point anywhere on the sky with an open 
loop accuracy of $<$10''. Thus, source acquisition is straightforward,
and easily automated.
The CCD camera has liberated astronomers from the drudgery
of the darkroom, and the anguish of the interpretation of non-linear
photographic media. The CCD based camera produces digital data with
linear response, and with a quantum efficiency as high as \ap90\%,
two orders of magnitude better than photographic film. Also,
CCDs can provide simultaneous measures
of sky brightness and comparison star brightness, which
permits accurate differential photometry even with a partly cloudy
sky.

A number of groups are operating automatic telescopes and some are
developing  plans for networks of automatic
telescopes (hypertext links to those with web pages are maintained at
http://gamma.bu.edu/atn/auto\_tel.html).
At least three manufacturers have designed telescopes of aperture 60 cm
or larger which are capable of automated operation, Torus Technologies 
and DFM in this country, and TTL in England.

A coordinated network of automated telescopes at diverse sites
 will facilitate optical monitoring of 
selected GLAST sources on sub-day timescales,
a task which is not otherwise routinely feasible --- although we have 
done this experimentally with a miniscule duty-cycle
with the 
Whole Earth Blazar Telescope (WEBT)
which is described in a separate paper in this volume
(Mattox, 1999b). The ATN project has been undertaken to
promote the development of a network of automated telescopes to
support blazar monitoring. A
web site has been established at http://gamma.bu.edu/atn/
to coordinate effort, and disseminate
information.

\vskip .5cm
{\bf 4. Automatic Telescope Standards}
\vskip .2cm

Much remains to be done in the realm of
software before automatic telescopes can execute
a program such as the GLAST blazar monitoring.
There are currently no standards in place for automated telescopes, to
permit them to be used coherently.

Therefore, an international working group is in formation 
to work with the IAU Commission (number 9) on Instruments to
develop
standards for automatic telescopes. The  web site is
http://gamma.bu.edu/atn/standards/.
These standards will
expedite the creation and utilization of networks of telescopes for
science and education. 

The existence of a standard command set will form an interface between a 
telescope specific TCS (Telescope Control System) and a higher 
level Observatory Control System (OCS). This will promote the
development of telescope-independent OCS software, which
will provide for instant robotocization of additional
new and refurbished telescopes which comply with the TCS
standard.

The standards will also include appropriate protocol for 
Internet control of automatic telescopes. The existence of such a
standard protocol will promote cooperative development and utilization
of networks of robotic telescopes. 
A standard protocol
for Internet control will facilitate cooperative utilization
of these telescopes. This will provide for the utilization of
more diverse facilities by all participants, increasing the
range of projects possible, and the efficiency of telescope
utilization.

\vskip .5cm
{\bf 5. Other Scientific Applications of the Networks of 
Automatic Telescopes}
\vskip .3cm

It is also expected 
that  networks of 
automatic telescopes  will be
useful for studying other transient 
phenomena, e.g., 
binary stellar systems,
gamma-ray bursts,
quasar/galaxy lensing systems,
microlensing events, and
asteroseismology. 
A network which is sized to provide observing time for other areas 
of investigation 
will include more telescopes. Therefore, it will be more efficiently 
scheduled,
and will provide better multi-longitude coverage for blazars, and 
gamma-ray burst
follow-up. Therefore, it is of interest to collaborate with
other astronomers who can use automatic optical telescopes.

Automatic telescopes can
also serve as a very 
valuable facility for science education. 
A network of automatic telescopes is being proposed by the
Hands-On Universe Project
(http://hou.lbl.gov/)
to  provide abundant, high-quality, CCD data for education.
The possibility of integrating blazar monitoring into this
effort is being explored.

{\bf References}

Bloom, S.D. \& Marscher, A.P., 1993, CGRO Sym,  AIP Conf. Proc. \#280, p. 578

Dermer, C. D., Schlickeiser, R., \& Mastichiadis, A., 1992, A\&A 256, L27

Ghisellini, G., \& Madau, P., 1996, MNRAS, 280, 67

Hartman 1999,  \etal, ``3rd EGRET Catalog'', ApJS, in press

Hudson, K.I., Chiu, H.Y., Maran, S.P., Stuart, F.E., Vokac, P.R., 1971, ApJ, 165, 573

Maran, S.P., 1967, Science, 158, 867

Mattox, J.R., 1999a, ``The Identification of 3EG EGRET Sources with
Flat-Spectrum Radio Sources'', ApJS, in preparation

Mattox, J.R., 1999b, ``The Whole Earth Blazar (WEB) Telescope'', this volume

Mattox, J.R., \etal, 1997, ApJ, 476, 692

Sikora, M., Begelman, M. C., \&  Rees, M. J. 1994, ApJ 421, 153

Stecker, F.W. and Salamon, M.H., 1996, ApJ, 464, 600

Takahashi, T., \etal, 1996, ApJ, 470, L89

Wehrle, A.E., \etal, 1998, ApJ, 497, 178 

\end{document}